\documentclass[journal]{IEEEtran}
\ifCLASSINFOpdf
\else
\fi
\usepackage{cite}
\usepackage{graphicx}
\usepackage{amssymb}
\usepackage{amsmath}
\usepackage{multirow}
\usepackage{booktabs}

\begin{document}
%
\title{A Weakly Supervised Segmentation Network Embedding Cross-scale Attention Guidance and Noise-sensitive Constraint for Detecting Tertiary Lymphoid Structures of Pancreatic Tumors}
%
%
%
\author{Bingxue Wang, Liwen Zou, Jun Chen, Yingying Cao, Zhenghua Cai, Yudong Qiu, Liang Mao, Zhongqiu Wang, Jingya Chen, Luying Gui and Xiaoping Yang
\thanks{Bingxue Wang and Luying Gui are with School of Mathematics and Statistics, Nanjing
University of Science and Technology, Nanjing, 210094, P. R. China.}
\thanks{Liwen Zou and Xiaoping Yang are with Department of Mathematics, Nanjing University, Nanjing, 210093, P. R. China.}
\thanks{Jun Chen is with Department of Pathology, Nanjing Drum Tower Hospital, Nanjing, 210008, P. R. China.}
\thanks{Yingying Cao, Zhongqiu Wang and Jingya Chen are with Department of Radiology, Affiliated Hospital of Nanjing University of Chinese Medicine, Nanjing, 210000, P. R. China.}
\thanks{Zhenghua Cai is with Medical School, Nanjing University, Nanjing, 210007, P. R. China.}
\thanks{Yudong Qiu and Liang Mao  are with Department of General Surgery, Nanjing Drum Tower Hospital, Nanjing, 210008, P. R. China.}
\thanks{Corresponding authors: Xiaoping Yang (e-mail: xpyang@nju.edu.cn) and Luying Gui (e-mail: ly.gui@njust.edu.cn)}
}


%
%

\markboth{IEEE JOURNAL OF BIOMEDICAL AND HEALTH INFORMATICS, VOL. XX, NO. XX, XXXX 2023}%
{Wang \MakeLowercase{\textit{et al.}}: A Weakly Supervised Segmentation Network Embedding Cross-scale Attention Guidance and Noise-sensitive Constraint for Detecting Tertiary Lymphoid Structures of Pancreatic Tumors}
%



\maketitle

\begin{abstract}
The presence of tertiary lymphoid structures (TLSs) on pancreatic pathological images is an important prognostic indicator of pancreatic tumors. Therefore, TLSs detection on pancreatic pathological images plays a crucial role in diagnosis and treatment for patients with pancreatic tumors. However, fully supervised detection algorithms based on deep learning usually require a large number of manual annotations, which is time-consuming and labor-intensive. In this paper, we aim to detect the TLSs in a manner of few-shot learning by proposing a weakly supervised segmentation network. We firstly obtain the lymphocyte density maps by combining a pretrained model for nuclei segmentation and a domain adversarial network for lymphocyte nuclei recognition. Then, we establish a cross-scale attention guidance mechanism by jointly learning the coarse-scale features from the original histopathology images and fine-scale features from our designed lymphocyte density attention. A noise-sensitive constraint is introduced by an embedding signed distance function loss in the training procedure to reduce tiny prediction errors. Experimental results on two collected datasets demonstrate that our proposed method significantly outperforms the state-of-the-art segmentation-based algorithms in terms of TLSs detection accuracy. Additionally, we apply our method to study the congruent relationship between the density of TLSs and peripancreatic vascular invasion and obtain some clinically statistical results.
\end{abstract}

\begin{IEEEkeywords}
Tertiary lymphoid structures, Pancreatic tumor, Weakly supervised segmentation, Cross-scale attention, Noise-sensitive constraint.
\end{IEEEkeywords}

%
\IEEEpeerreviewmaketitle

\section{Introduction}
%
%
%
%
\IEEEPARstart{H}{istopathology} images are often the gold standard for disease detection, diagnosis, and prognostic analysis. However, analysis for such images is a very challenging task because of their large sizes and numerous elements. In recent years, the development of deep neural networks has led to many breakthroughs in automatic histopathology image classification and segmentation \cite{madabhushi2016image, kiemen2022coda}. These methods depend highly on extensive training and accurate pixel-level labels, which are particularly labor-intensive and time-consuming due to the huge sizes of pathological images. Actually, only a trained pathologist can distinguish precisely among the elements. Therefore, acquiring accurate annotations on pathological images is more complicated than annotating other medical images. The development of weakly supervised and unsupervised learning methods has become an inevitable trend in pathological image processing \cite{zhang2021joint, xu2019camel}.

In pathology research, the immune micro-environment is a topic of great concern. The main research target of the immune micro-environment is lymphocytes. Their particular aggregation clusters, known as tertiary lymphoid structures (TLSs), have received significant attention in recent years.
Actually TLSs are discrete structured tissues of infiltrating immune cells, or in other words, more organized aggregation structures of lymphocytes. In some studies, TLSs are considered as any lymphoid aggregate similar to secondary lymphoid organs in non-lymphoid structures \cite{munoz2020tertiary}.
Several researches \cite{sautes2019tertiary,schumacher2022tertiary} concentrated on the presence, composition, and location of TLSs in health and disease tissues and organs. The high density of TLSs found in many cancers is associated with prolonged survival of patients, such as colorectal cancers \cite{trajkovski2018tertiary,bergomas2011tertiary} and breast cancers \cite{heindl2018relevance}.

Previous studies have generally considered pancreatic tumors as cold tumors and lack of immune response \cite{mortezaee2021enriched, redding2023splendid}. In contrast, some studies in recent years have found that the immune micro-environment of pancreatic tumors is highly heterogeneous, which largely contributes to its lack of mechanical clarity to date. One of the critical elements of its heterogeneity is a large number of immune cell subsets \cite{rubin2022tumor}. This means accurately segmenting lymphocytes and classifying clusters are crucial for studying pancreatic tumors.  In the pathologist's opinion, lymphocytes in a TLS have a particular specific aggregation pattern, such as having a distinctive high density. Although a TLS does not have an envelope but is still clearly delimited from the surrounding tissues. However, it is challenging to identify all TLSs on a pathological image because of the considerable size variants of TLSs and their irregular distributions on the histopathology images. In addition, backgrounds with different colors and textures may interfere with the recognition of TLSs by humans.
Also, an essential feature of the pancreas lacking immune response determines that there are fewer lymphocytes and TLSs in the pancreas, and it is more difficult to obtain a large number of training samples.

In order to accurately detect TLSs on
a small-scale dataset containing whole slide images (WSIs) of hematoxylin and eosin (H\&E) stained histopathology, we propose a weakly supervised segmentation network embedding cross-scale attention guidance and noise-sensitive constraint in this work. The proposed method is mainly divided into three parts: (1) all lymphocyte nuclei are segmented and identified from the unlabeled pancreatic pathology images by combining a pretrained nuclei segmentation model and adversarial learning. The corresponding lymphocyte density maps are constructed; (2) based on the aggregation characteristics of TLSs, we convert the lymphocyte density map in WSI to the attention guidance and establish a cross-scale attention network to learn the TLS features from different scales; (3) considering that the location of a TLS is critical information, we introduce a noise-sensitive constraint by a signed distance function to train a segmentation network with weak bounding box annotation. It is designed to make the network explicitly learn the TLSs' location distribution. To verify the effectiveness of our method in few-shot learning, the proposed method is trained on a small-scale dataset and evaluate on another large-scale dataset. The experimental results show that our method can accurately detect TLSs, and outperforms the state-of-the-art (SOTA) segmentation-based algorithms in terms of detection accuracy. As a clinical application, we use the density of TLSs to predict peripancreatic vascular invasion based on our proposed method.

The major contributions of this work are listed as follows.
\begin{itemize}
\item To the best of our knowledge, it is the first work to realize the few-shot detection task of TLSs on pancreatic histopathology images by embedding cross-scale attention guidance and noise-sensitive constraint into a weakly supervised segmentation network, which can be applied to the TLS analysis for other tumors.

\item We construct the density map of lymphocytes which reflects the distribution characteristics of TLSs on pathological images and propose a cross-scale attention mechanism by jointly learning the TLS features from the coarse-scale pathological images and the fine-scale lymphocyte density maps. 

\item A noise-sensitive constraint with a signed distance function loss is introduced for training the TLS segmentation network with weak bounding box annotations, which helps to explicitly learn the TLS distribution and avoid enormous performance drops caused by tiny predicted errors. 

\item Experimental results show that our method significantly outperforms the state-of-the-art (SOTA) segmentation-based algorithms in terms of the TLS detection accuracy on pancreatic pathological images. And we validate that the TLS density is significantly related to the peripancreatic vascular invasion by our proposed method based on the clinical data acquired from two independent institutions.

\end{itemize}

\section{Method}\label{sec2}
\begin{figure*}[htbp]%
\centering
\includegraphics[width=15cm]{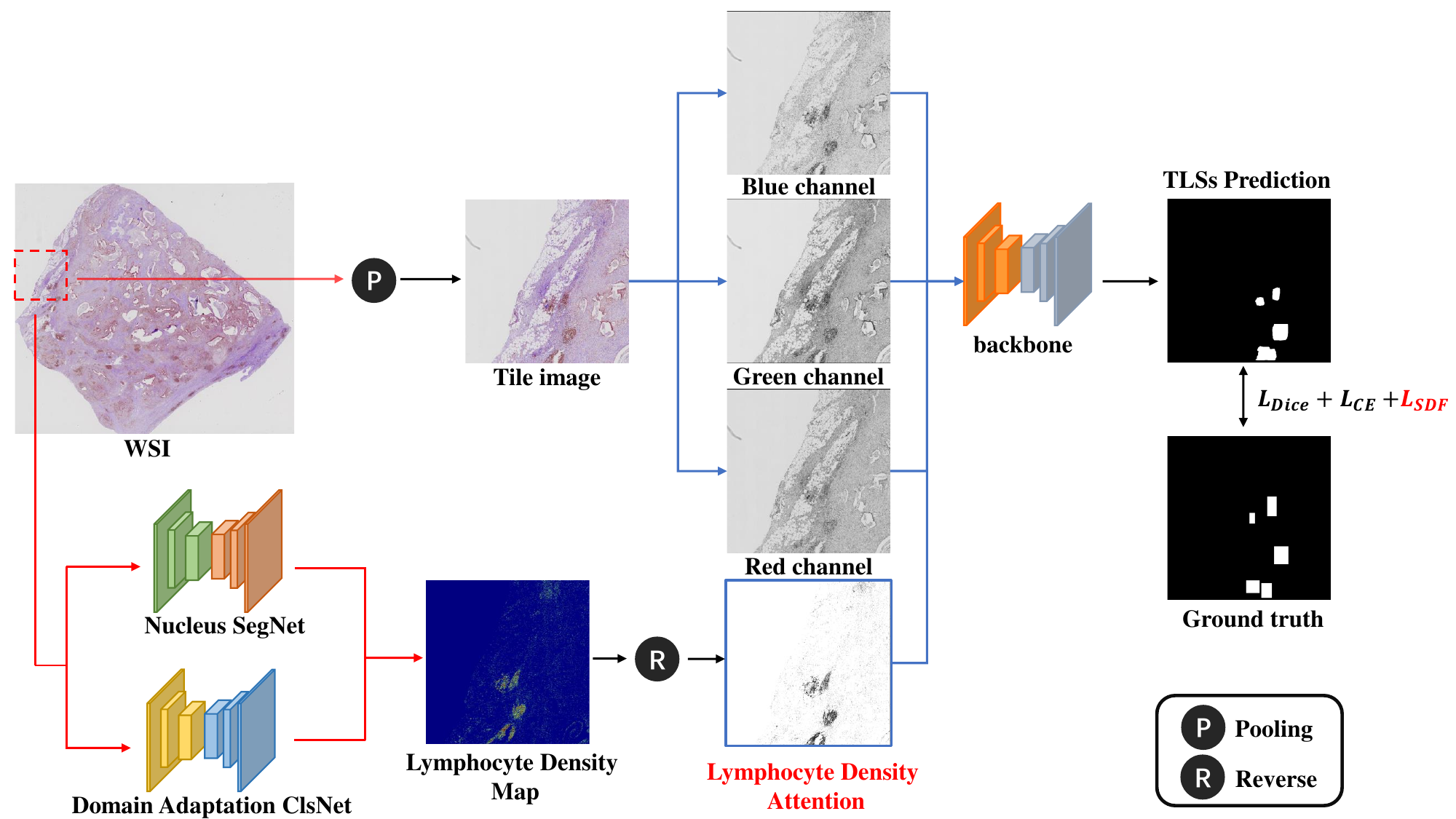}
\caption{Pipeline of our proposed segmentation network embedding cross-scale attention guidance and noise-sensitive constraint for weakly supervised TLSs segmentation of pancreatic cancer.}\label{fig1}
\end{figure*}

We propose a weakly supervised segmentation network for the few-shot detection of TLSs on pancreatic pathological images by embedding cross-scale attention guidance and noise-sensitive constraint to improve the TLS segmentation performance. The pipeline of our proposed method is shown in Figure \ref{fig1}.  There are three core  modules in our proposed model: (1) Lymphocyte nuclei segmentation and classification by combining nuclei segmentation baseline model and domain adversarial learning; (2) cross-scale attention guidance mechanism by jointly learning the TLS features from the coarse-scale H\&E image and the fine-scale lymphocyte density map; (3) a noise-sensitive constraint by embedding a sign distance function loss for training the segmentation network with weak bounding box annotations. 

When the proposed model is fed with an H\&E WSI from a patient with  pancreatic tumor, there are two processing flows for extracting cross-scale features of the TLS targets. The original H\&E image is fed into a pretrained baseline model for nuclei segmentation and a domain adversarial network for lymphocyte nuclei classification. Then, we construct the lymphocyte density map as the fine-scale feature on each $d \times d$ tile image based on the previous predictions. Meanwhile, i order to learn the generic global context information, we get the coarse-scale feature from the original image by $d \times d$ mean pooling operation to make it the same size as the lymphocyte density map. Additionally, we utilize a reverse operation on the lymphocyte density map to make it be an attention map, termed as the lymphocyte density attention (LDA), which is compatible with the original H\&E image in terms of the intensity distribution. Then, a cross-scale attention guidance mechanism based on a U-shape backbone with four-channel inputs is proposed to process the above cross-scale images. It learns the macroscopic features from the coarse-scale H\&E image and microscopic features from the fine-scale lymphocyte density attention. Additionally, the proposed segmentation model is trained with bounding box annotations, which is weakly supervised for the segmentation task. Furthermore, a noise-sensitive constraint with a signed distance function loss (SDF) is used in the training procedure to explicitly learn the TLS distribution and avoid huge performance drops caused by tiny predicted errors.

\subsection{Segmentation and classification for lymphocyte nuclei}\label{L-seg}
The immune response in pancreatic tumors is usually not strong, and the distribution of immune cells is relatively sparse. In contrast, we note that the lymphocyte nuclei' morphological characteristics are independent of the organs in which they are located. The lymphocyte nuclei on pathology images of other organs should be morphologically identical to the pancreatic lymphocyte nuclei, although these images have different backgrounds. Therefore, we considered 
transferring the lymphocyte nuclei from a public dataset with lymphocyte nuclei annotations of other organs to our pancreatic pathological dataset without lymphocyte nuclei annotations.

\begin{figure}[htbp]
		\centering
		{
			\includegraphics[width=8.5cm,]{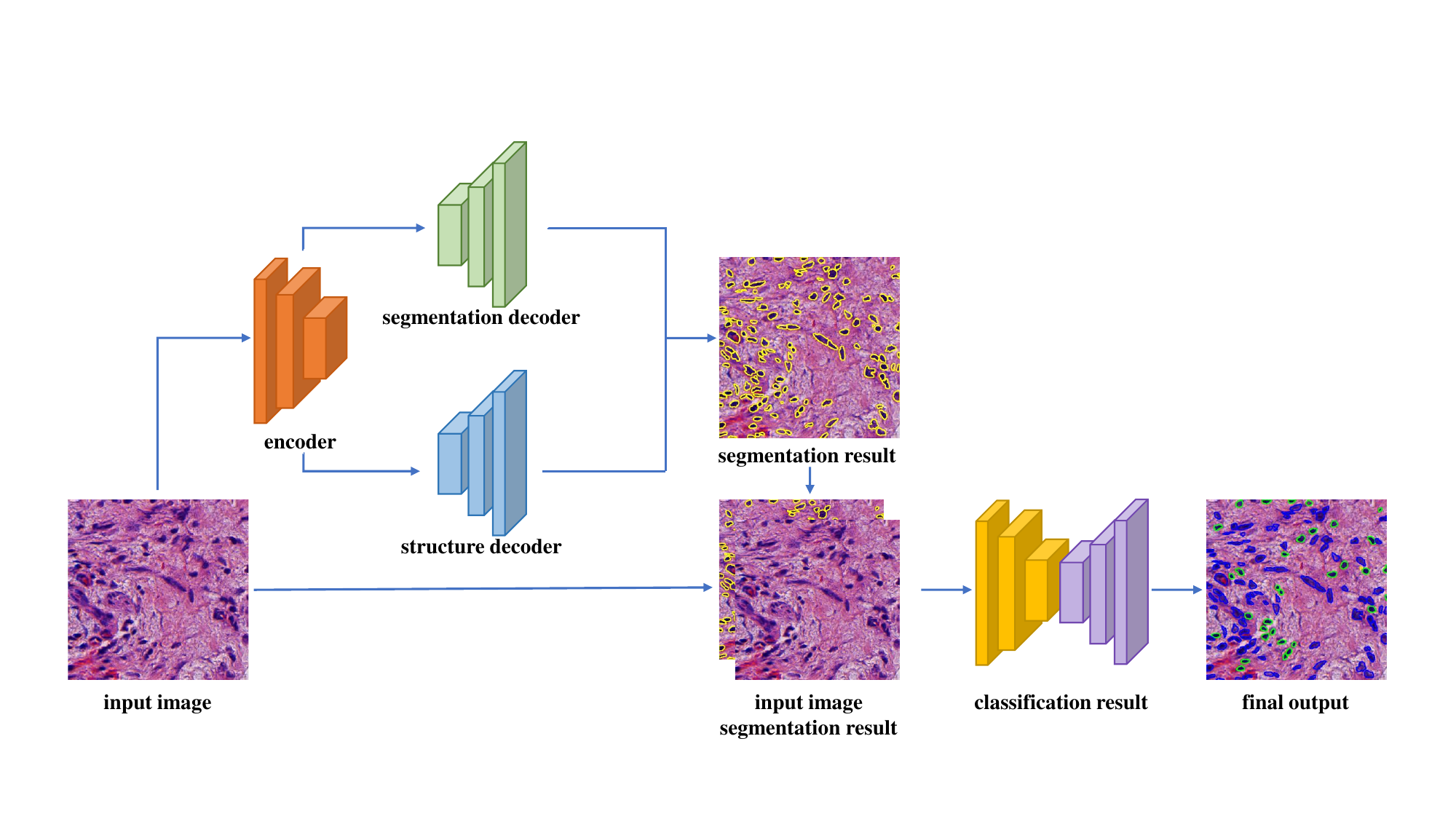}
		}
		\caption{Structure of the designed combination of the nuclei segmentation baseline model and domain adversarial network for recognizing lymphocyte nuclei.}\label{HD}
	\end{figure}

\begin{figure}[htbp]
		\centering
		{
			\includegraphics[width=8.5cm,]{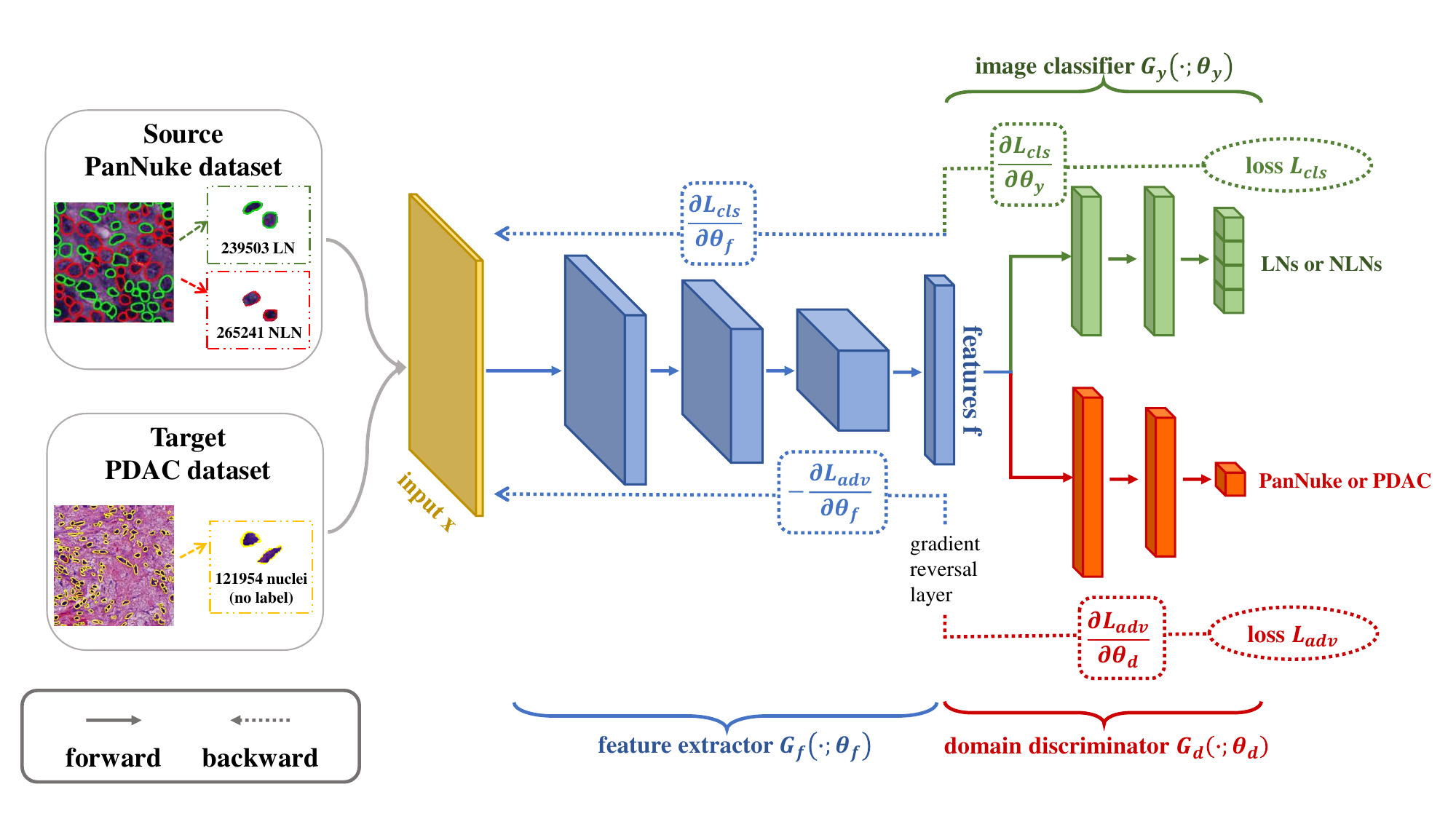}
		}
		\caption{The domain adversarial neural network training strategy based on the annotated lymphocyte nuclei dataset from PanNuke and our nuclei dataset obtained from the pretrained HoVer-Net model. LN and NLN denote lymphocyte nuclei and non-lymphocyte nuclei, respectively.}\label{dann}
	\end{figure}

To this end, we introduce the combination of a pretained baseline model for nuclei segmentation and a domain adversarial network for lymphocyte nuclei classification shown in Figure \ref{HD}. HoVer-Net is a robust baseline model for nuclei segmentation and has demonstrated state-of-the-art performance on various nuclei segmentation tasks \cite{2019HoVer}. There are three branches in the HoVer-Net architecture: the segmentation branch for getting coarse prediction, the structure branch for solving the nuclei overlapping, and the classification branch for nuclei recognition. We use the segmentation and structure branches from the pretrained HoVer-Net in the PanNuke dataset  \cite{gamper2020pannuke} to segment the nuclei for our task. As for the classification, we train a domain adversarial neural network (DANN)  \cite{2015Domain} to get better performance on lymphocyte nuclei recognition. A domain adversarial training strategy is used in the classification network. Figure \ref{dann} demonstrates our training strategy for domain adversarial learning. We crop 239503 lymphocyte nuclei images and 265241 non-lymphocyte nuclei images from the PanNuke dataset with nuclei annotations. Then we obtain 121954 nuclei images with unknown categories (lymphocyte or non-lymphocyte nuclei) based on the segmentation results from the pretrained HoVer-Net. We choose the former as the source domain and the latter as the target domain to train the domain adversarial network for lymphocyte nuclei recognition. ResNet18 \cite{2016Deep} is used as the feature extractor, and two loss terms are calculated as follows.

\begin{equation}\label{losscls}
  L_{cls}=\sum_{i=1}^M y_i\log\frac{1}{G_{y}(G_{f}(x_i))},
\end{equation}

\begin{equation}\label{lossadv}
  L_{adv}=\sum_{i=1}^M d_{i}\log\frac{1}{G_{d}(G_{f}(x_{i}))}+
  (1-d_{i})\log\frac{1}{G_{d}(G_{f}(x_{i}))},
\end{equation}
where $G_f$, $G_y$, and $G_d$ denote the feature extractor, image classifier, and domain discriminator, respectively. $M$ denotes the number of training samples. $x_i$, $y_i$, and $d_i$ denote the image input, nucleus category annotation, and domain label of sample $i$, respectively.
The overall loss of DANN is defined by 
\begin{equation}\label{loss}
  L_{DANN}=L_{cls}+L_{adv}
\end{equation}

\subsection{Cross-scale attention guidance mechanism}
\begin{figure}[htbp]
		\centering
		{
			\includegraphics[width=8.5cm,]{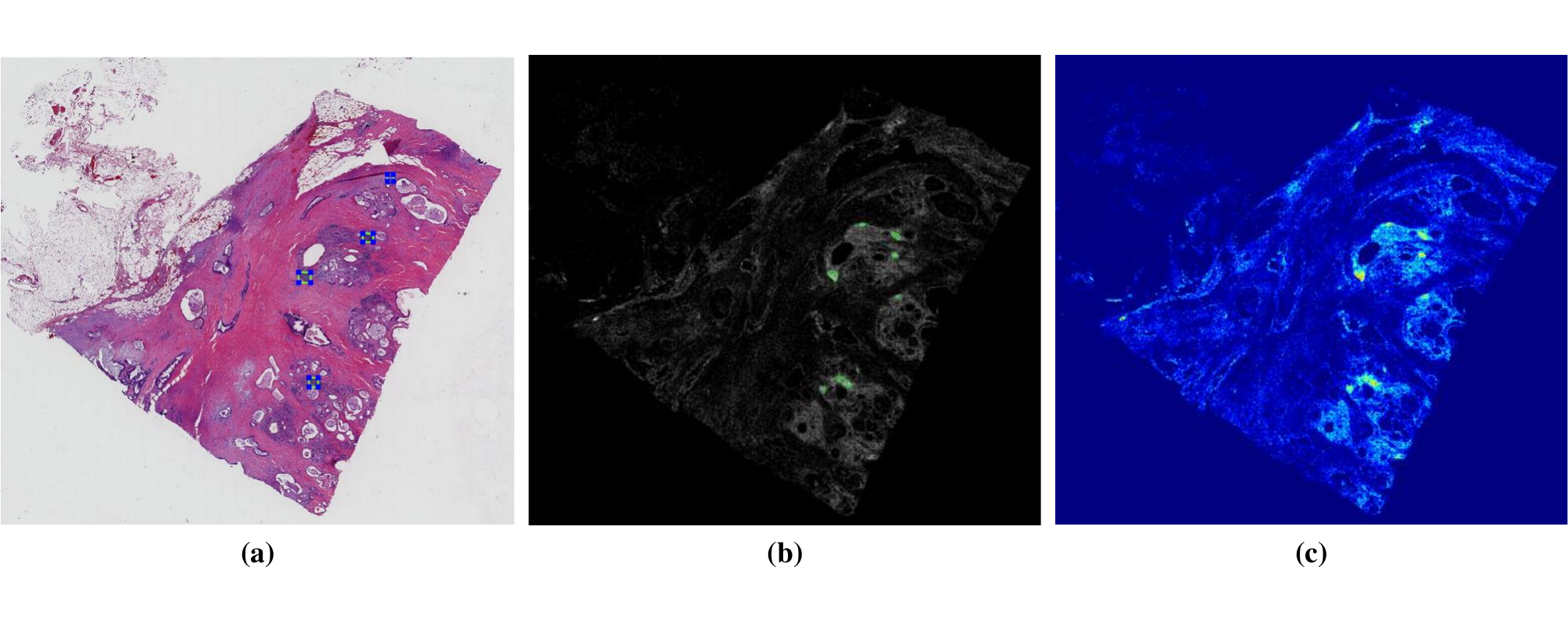}
		}
		\caption{The visualization of the calculated lymphocyte density map of a WSI from our collected dataset. (a) WSI with TLS annotations (shown in blue boxes); (b) Gray-scale image of the calculated lymphocyte nuclei density map, high density regions are denoted by green contours; (c) Heatmap of the calculated lymphocyte nuclei density.}\label{ldm}
	\end{figure}

TLSs are the particular forms of lymphocyte aggregations, and it is natural to look for TLSs where lymphocytes are relatively aggregated. Therefore, we calculate the lymphocyte density map based on the detected lymphocyte nuclei and convert it to be attention map to guide the detection of TLSs.
To obtain the lymphocyte density map on a WSI, we count the number $N_{ij}$ of predicted lymphocyte nuclei in each non-overlapping $d \times d$ patch at location $(i,j)$. Each pixel occupies $25 \times 25$ $\mu m^2$. Then we get a gray-scale lymphocyte density map by the following formulation:

\begin{equation}\label{loss}
  D_{ij}=255 \times \frac{N_{ij}-N_{min}}{N_{max}-N_{min}},
\end{equation}
where $N_{max}$ and $N_{min}$ are the maximum and minimum of the predicted number of lymphocyte nuclei among the patches. Figure \ref{ldm} shows the lymphocyte density map calculated for a WSI from our collected dataset. We split the three channels of the original H\&E images into blue, green, and red channels shown in Figure \ref{fig1}. It can be found that the target TLS regions show low intensity in each color channel. We add a reverse operation to make the predicted lymphocyte density map compatible with the original H\&E image to generate our desired attention map. Each lymphocyte density attention $A_{ij}$ at location $(i,j)$ is calculated as $A_{ij}=255-D_{ij}$. Then a cross-scale attention guidance network is established on the U-shape backbone architecture. There are four channel inputs of the backbone to jointly learn the coarse-scale features from the original H\&E images and the fine-scale features from the calculated lymphocyte density attention.

\subsection{SDF loss for noise-sensitive constraint}
\begin{figure}[htbp]
		\centering
		{
			\includegraphics[width=8.5cm,]{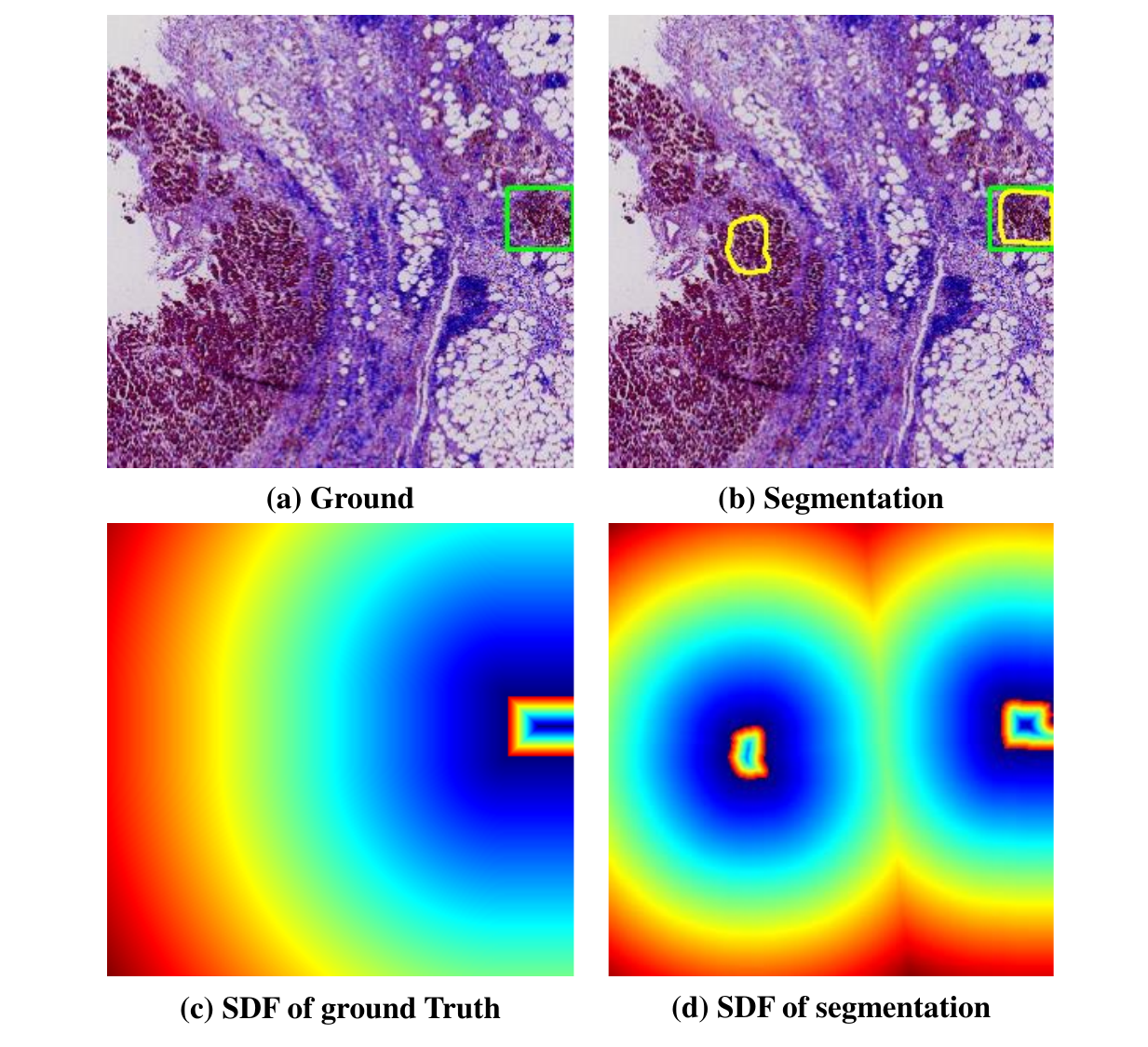}
		}
		\caption{Visual comparison between the SDF of ground truth and SDF of segmentation.  (a) and (b) are the ground truth (green rectangle) and the segmentation (yellow contours) of TLSs. (c) and (d) are the corresponding signed distance function representation. It can be found that tiny segmentation errors can significantly affect the SDF distribution.}\label{sdf}
	\end{figure}

Considering that accurate positioning usually has high priority than finding the boundary of a TLS for clinical issues. Therefore, we train the proposed network with the bounding box annotations from the experts, which is weakly supervised for a segmentation task. Generally, it is difficult for a semantic segmentation network to detect small objects by training with general segmentation loss, such as Dice loss and cross-entropy (CE) loss, because these small regions almost have no impact on the overall loss. However, minor prediction errors can lead to misdiagnosis. To solve this problem, we introduce a noise-sensitive constraint by embedding a signed distance function (SDF) loss into the overall loss function for the training procedure. The signed distance function of a predicted binary mask $Y$ can be calculated as follows:

\begin{equation}\label{dist}
	SDF(x,Y)=
	\left\{
	\begin{array}{lr}
    \underset{y \in \partial Y}{\min}-d(x,y), &x \in Y_{in} \\
	0, &x \in \partial Y \\
    \underset{y \in \partial Y}{\min}d(x,y), &x \in Y_{out},
	\end{array}
	\right.
	\end{equation}
where $d(x,y)$ is the Euclidian distance between $x$ and $y$, and $Y_{in}$, $Y_{out}$ and $\partial Y$ denote the inside, the outside and the boundary of the object, respectively. Then we get the signed distance loss by calculating the mean square error (MSE) between the SDFs of segmentation and ground truth:
\begin{equation}\label{loss}
  L_{SDF}(Y,\hat Y)=\frac{\sum_{i=1}^N(SDF(x_i,Y)-SDF(x_i,\hat Y))^2}{N}.
\end{equation}
where $N$ is the number of pixels in an input image. As shown in Figure \ref{sdf}, small segmentation errors significantly impact the SDF distribution. Besides, the traditional Dice and CE losses are also used in the training procedure. They can be calculated as follows:

\begin{equation}\label{loss}
  L_{Dice}=-\frac{2 \sum_{i=1}^N s_i g_i}{\sum_{i=1}^N s_i^2 + \sum_{i=1}^N g_i^2},
\end{equation}

\begin{equation}\label{loss}
  L_{CE}= -\frac{\sum_{i=1}^N g_i lnp_i}{N},
\end{equation}
where $s_i$ and $g_i$ denote the predicted segmentation and the ground truth of pixel $i$, respectively. $p_i$ denotes the softmax output of $s_i$. Therefore, the overall loss function of our proposed TLSs segmentation network can be formulated as follows:

\begin{equation}\label{loss}
  L=L_{Dice}+L_{CE}+L_{SDF}.
\end{equation}

\section{Experiments}
\subsection{Dataset}
In this work, we evaluate our proposed method on two datasets collected from Nanjing Drum Tower Hospital (NDTH) and Jiangsu Province Hospital of Chinese Medicine (JHCM), respectively. The NDTH  dataset consists of 38 WSIs from 12 surgical pathology-confirmed PDAC patients. The JHCM dataset is composed of 57 WSIs from 41 PDAC patients. To verify the generalization and the performance in few-shot learning of our proposed method, we train and validate our model on the smaller-scale NDTH dataset and test on the larger-scale JHCM dataset. This study was approved by the Ethics Committee of Nanjing Drum Tower Hospital and Jiangsu Province Hospital of Chinese Medicine.

The TLSs on the WSIs are annotated with bounding boxes by two experienced pathologists. The NDTH dataset is divided into four equal folds. One fold containing 9 WSIs is fixed as the validation set, and the other three folds containing 29 WSIs are adopted for training throughout the experiments. The JHCM dataset is only used for testing (i.e., zero-shot evaluation). 

\subsection{Evaluation metrics}
We employ the precision, recall and  $F_{\beta}$ score to measure the detection accuracy in our experiments, which can be calculated as follows:

\begin{equation}\label{pre}
  Precision=\frac{TP}{TP+FP},
\end{equation}

\begin{equation}\label{rec}
  Recall=\frac{TP}{TP+FN},
\end{equation}

\begin{equation}\label{fscore}
  F_{\beta}=(1+{\beta}^2) \times \frac{Precision\times Recall}{{\beta}^2 \times Precision+Recall},
\end{equation}
where $TP$, $FP$ and $FN$ denote the true positive, false positive and false negative TLS predictions, respectively. Generally, $F_1$ score (i.e., $\beta=1$ in Eq. \ref{fscore}) is used to be the harmonic mean of both the precision and recall. However, because the false negative predictions generally bring greater harm than the false positive predictions in clinical issues, which means the recall metric is more important than the precision metric. Therefore, we also use the $F_2$ score (i.e., $\beta=2$ in Eq. \ref{fscore}) to evaluate the detection performance with greater weight on recall.

\begin{figure}[htbp]
		\centering
		{
			\includegraphics[width=7.5cm,]{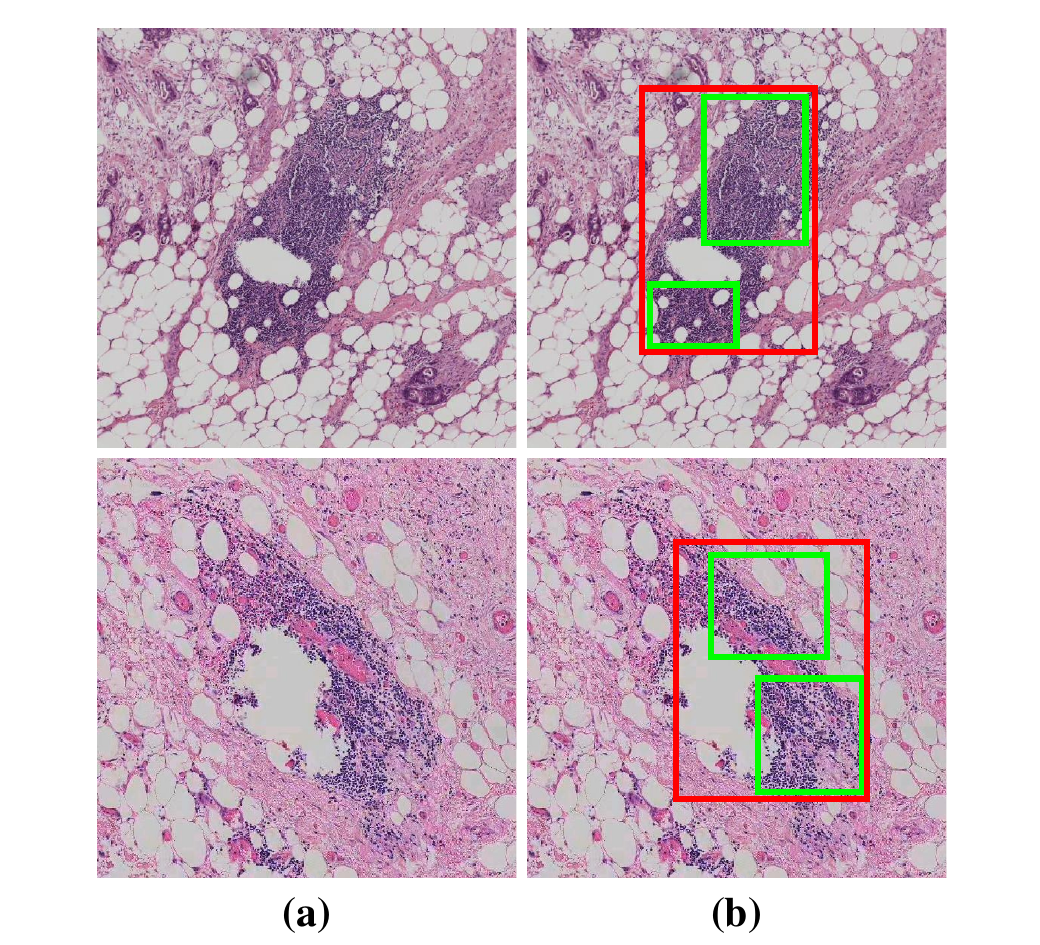}
		}
		\caption{Two examples to illustrate the annotation difference by different experts. (a) Image; (b) The TLS annotations by two experts which are denoted by red and green rectangles, respectively.}\label{andiff}
	\end{figure}

Moreover, because of the difference between the TLS annotations by different experts, especially in the regions where it is difficult to distinguish whether there is a single TLS or aggregated TLSs. Fig. \ref{andiff} shows two examples to describe the annotation difference.

\begin{figure}[htbp]
		\centering
		{
			\includegraphics[width=8.5cm,]{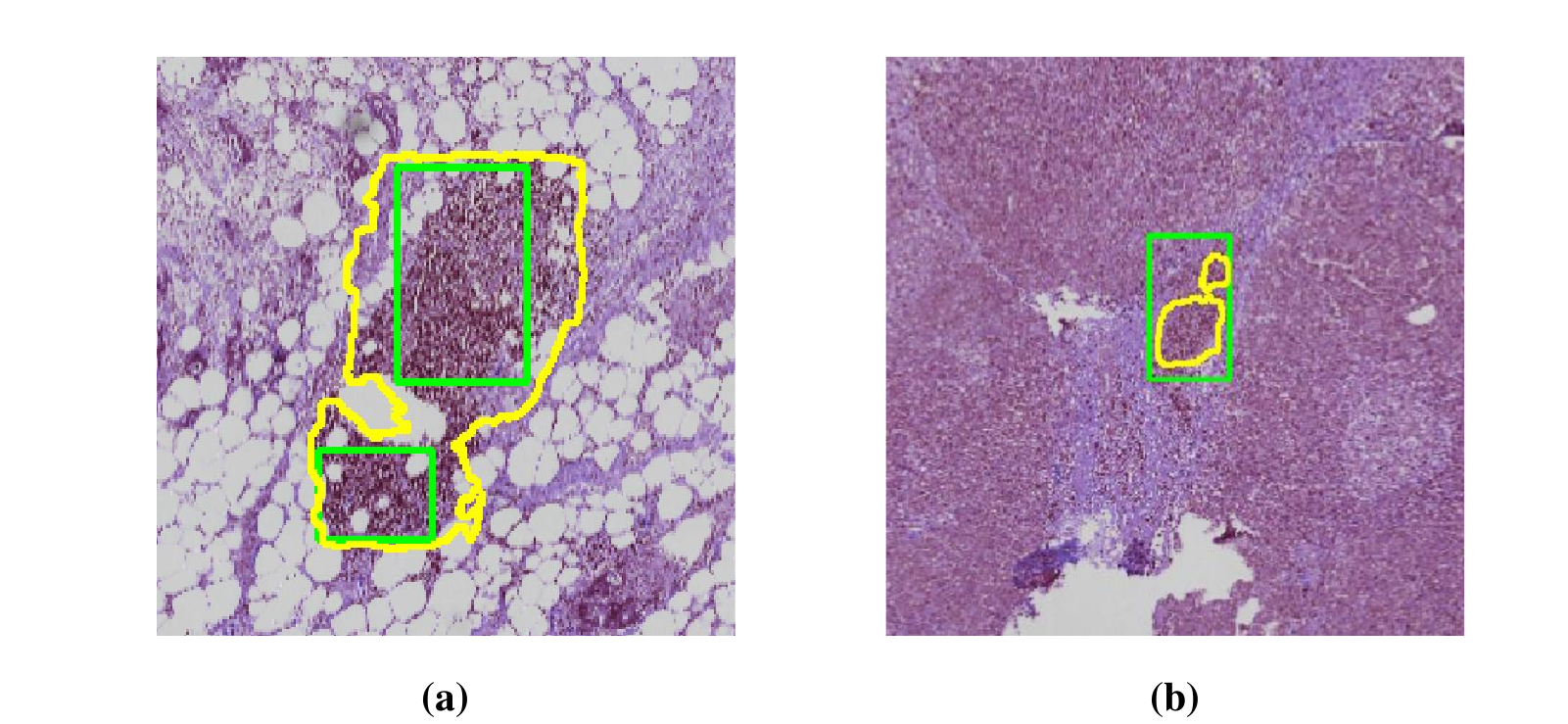}
		}
		\caption{Two examples to illustrate the proposed evaluation metrics.(a) When one predicted segmentation denoted by yellow contour covers two annotated boxes denoted by green rectangles, there are $TPS=1$ and $TPB=2$. (b) When one annotated box denoted by green rectangle is covered by two predicted segmentation denoted by yellow contours, there are $TPS=2$ and $TPB=1$.}\label{metric}
	\end{figure}

Therefore, we propose some new evaluation metrics to coordinate the annotation difference between the experts. Without loss of generality, since the precision describes how many positive cases are true among all the model predictions, we introduce the $TPS$ here instead of $TP$ to represent the TLS number of segmentation results that overlap with ground truths. Similarly, the recall describes how many annotated boxes are truly predicted, we introduce the $TPB$ instead of $TP$ to represent the number of annotated bounding boxes that overlap with the segmentation results. It can be seen that the $TPS$ and $TPB$ are generalizations of the general $TP$ without the one-on-one constraint. Then, we define the corresponding segmentation precision (SP) and box recall (BR) as follows:

\begin{equation}\label{sp}
  SP=\frac{TPS}{TPS+FP},
\end{equation}

\begin{equation}\label{br}
  BR=\frac{TPB}{TPB+FN}.
\end{equation}
Similarly, the general $F_{\beta}$ score ($GF_{\beta}$) can be calculated as follows:
\begin{equation}\label{fscore}
  GF_{\beta}=(1+{\beta}^2) \times \frac{SP\times BR}{{\beta}^2 \times SP + BR}.
\end{equation}
Fig. \ref{metric} illustrates the insight of the proposed metrics. 

\subsection{Ablation study}
In this section, we present the ablation study to verify the effectiveness of our proposed LDA and SDF in our method. We show the ablation study results in Table \ref{ablations} to investigate the individual impact of the proposed LDA and SDF module.

\subsubsection{Effectiveness of LDA}
\begin{figure*}[htbp]
		\centering
		{
			\includegraphics[width=15cm,]{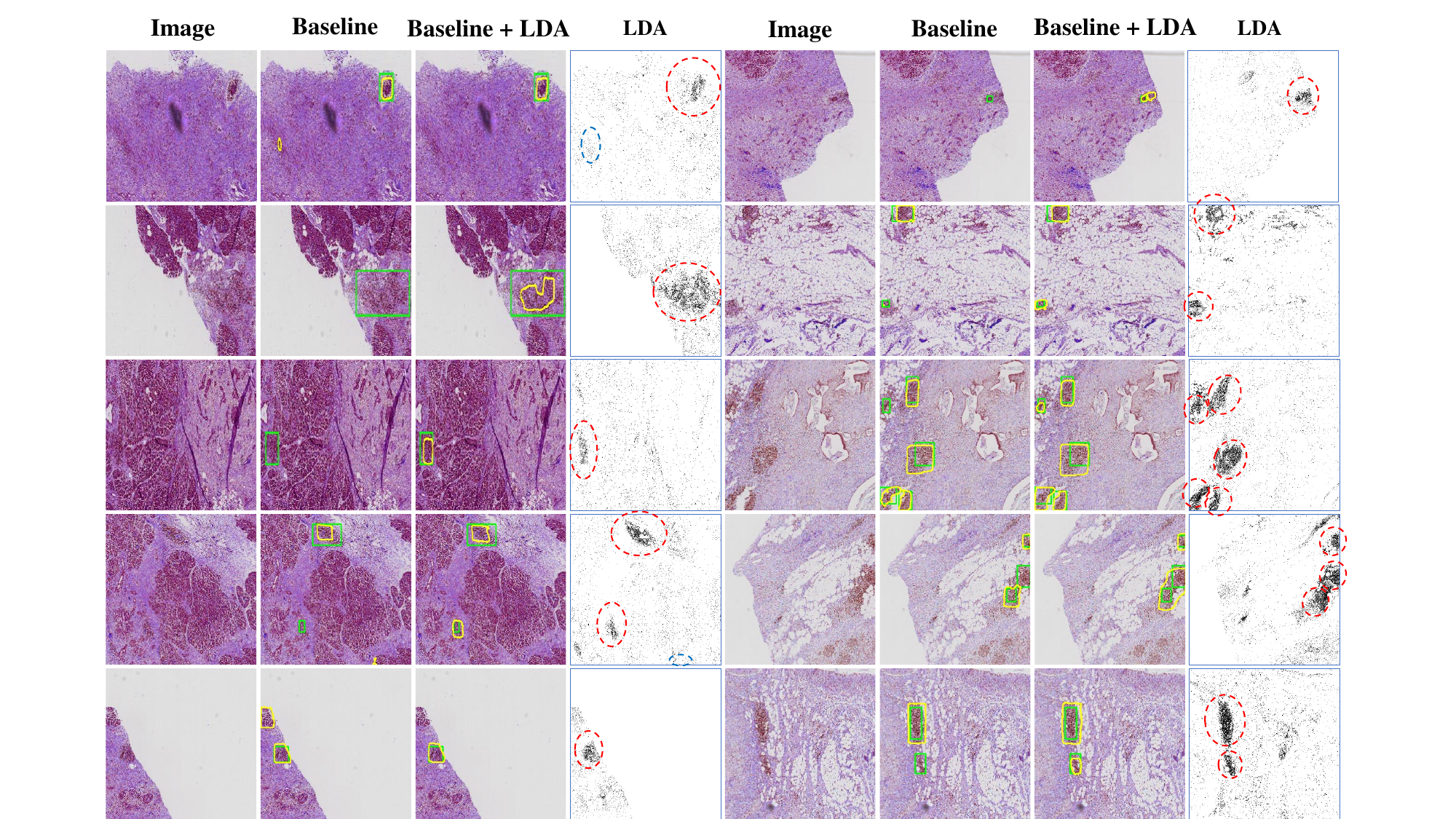}
		}
		\caption{Visual comparison of the ablation study for the proposed LDA. The baseline model is the proposed method without LDA and SDF loss. Regions with high and low attention on the calculated LDAs are signed with red and blue dashed circles, respectively. It can be found that the calculated LDA can significantly improve the TLS detection performance.}\label{LDA_vis}
	\end{figure*}

Without using LDA, the performance significantly drops on our collected datasets, leading to a decrease of 1.60\%-3.56\% in precision, 2.24\%-5.42\% in the recall, 3.16\%-3.42\% in $F_1$ score, and 2.70\%-4.60\% in $F_2$ score. For the introduced metrics, there is a performance decrease of 3.47\%-6.11\% in SP, 2.15\%-7.23\% in BR, 4.08\%-5.32\% in $GF_1$ score, and 3.01\%-6.46\% in $GF_2$ score without the LDA. Figure \ref{LDA_vis} shows the visual comparison of the ablation study for the proposed LDA. It can be found that the false positive predictions with low attention and false negative predictions with high attention can be efficiently captured.

\subsubsection{Effectiveness of SDF}
\begin{figure*}[htbp]
		\centering
		{
			\includegraphics[width=15cm,]{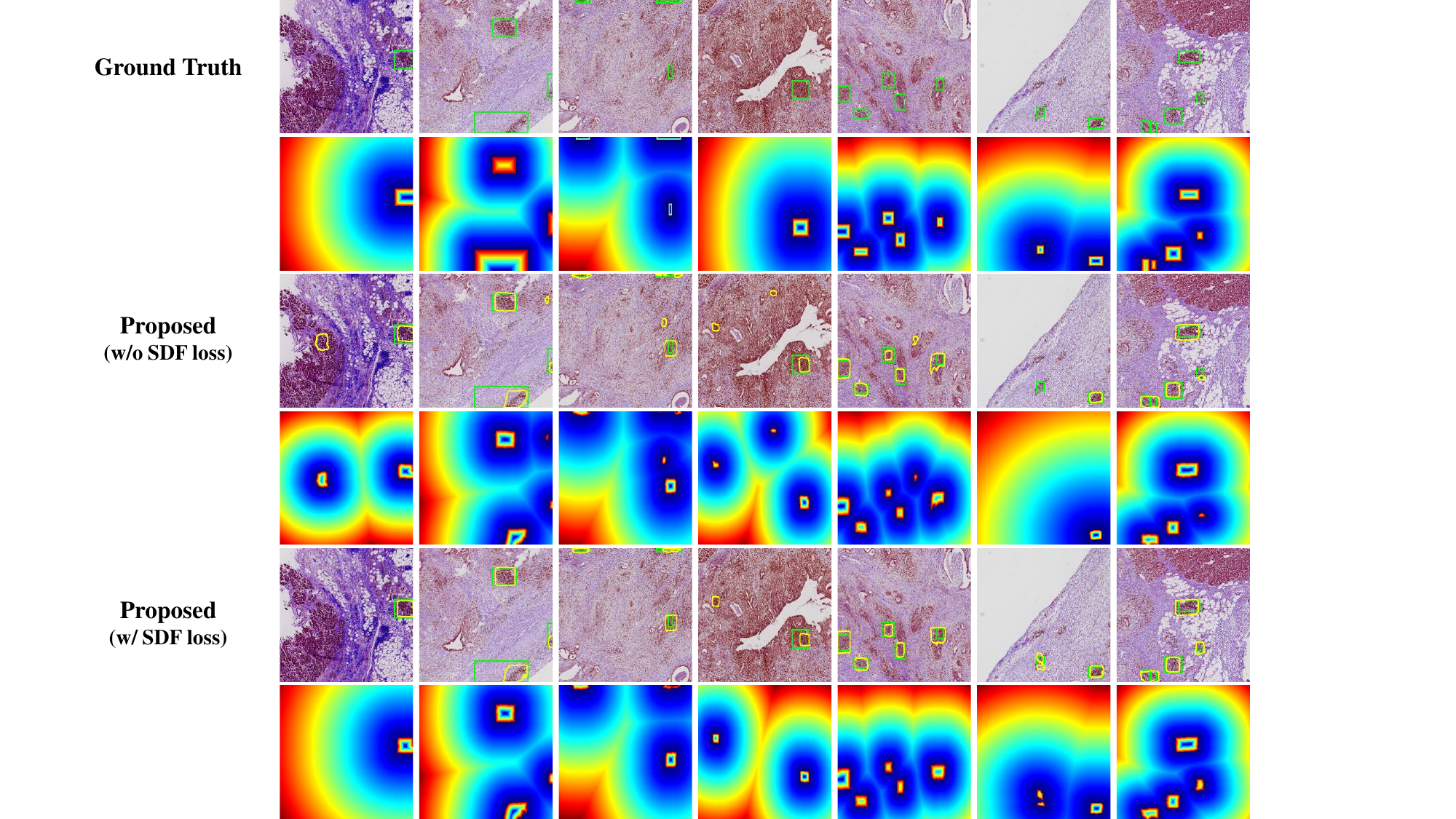}
		}
		\caption{Visual comparison of the ablation study for the SDF loss. Segmentation and ground truth are denoted by green and yellow contours in pathological images, respectively. The SDF maps are shown under the corresponding images.}\label{SDF_vis}
	\end{figure*}

Without using the SDF loss, the performance also drops on our collected  datasets, leading to a decrease of 2.39\%-5.48\% in precision, 1.85\%-3.45\% in $F_1$ score, and 1.27\%-1.48\% in $F_2$ score. For the introduced metrics, there is a performance decrease of 2.41\%-3.88\% in SP, 1.73\%-2.41\% in $GF_1$ score, and 0.51\%-2.41\% in $GF_2$ score without the SDF loss. It should be pointed that the SDF loss leads to a improvement of 1.21\% in recall on the NDTH dataset while it leads to a decrease of 0.84\% in recall on the JHCM dataset, which indicates that the SDF loss improves the overall performance of the detection model by mainly reducing the tiny false positive predictions instead of false negative predictions with relatively large sizes. Figure \ref{SDF_vis} presents the visual comparison of the ablation study for the SDF loss. We can see that tiny segmentation errors can cause huge losses in terms of the signed distance function map, and the SDF loss can significantly reduce the tiny false positive predictions.

\subsubsection{Effectiveness of LDA and SDF}Additionally, it can be observed that without using LDA and SDF losses leads to the worst performance in the recall, $F_1$ score, $F_2$ score BR, $GF_1$ score, $GF_2$ score, indicating that both the proposed LDA and SDF losses are essential for performance improvements. It should be pointed that the baseline model (the proposed method without LDA and SDF loss) achieves the best performance in segmentation precision on NDTH dataset, while the proposed method outperforms the baseline model by 0.29\% in the precision and achieves the best performance in all other metrics.

\begin{table*}[htbp]
\centering
\caption{Ablation study results of the proposed method on two collected datasets. Experimental results on the NDTH validation set and the JHCM dataset are presented. w/o means removing the corresponding module. P, R, F1 and F2 denote the precision, recall, $F_1$ score and $F_2$ score, respectively. SP, BR, GF1 and GF2 represent the introduced segmentation precision, box recall, $GF_1$ score and $GF_2$ score, respectively.}
\setlength{\tabcolsep}{4.2mm}{}{
\begin{tabular}{clcccccccc}
\toprule

Datasets                & Methods              & P      & R       & F1 & F2  &SP &BR &GF1  &GF2      \\ \midrule
\multirow{4}{*}{NDTH}   &Proposed       & \textbf{78.21}           & \textbf{84.34}      & \textbf{81.16}  &\textbf{83.04} &84.80 &\textbf{87.95} &\textbf{86.34} &\textbf{87.30}\\
                        &Proposed w/o $L_{SDF}$      & 75.82           & 83.13      & 79.31 &81.56  &82.39&85.54&83.93&84.89\\
				&Proposed w/o LDA        & 76.61          &78.92      & 77.74  & 78.44 &81.33&80.72&81.02&80.84\\
                &Proposed w/o LDA, $L_{SDF}$      & 77.92           & 72.29      & 75.00 & 73.35 &\textbf{86.21} &74.70&80.04&76.75 \\
                \midrule
\multirow{4}{*}{JHCM} &Proposed       & \textbf{62.25}           & 83.13      & \textbf{71.19}  & \textbf{77.90} & \textbf{76.15}           & 87.70      & \textbf{81.52}  & \textbf{85.12}  \\
                        &Proposed w/o $L_{SDF}$ & 56.77           & \textbf{83.97}     & 67.74  & 76.63  & 72.87           & \textbf{88.16}     & 79.79  & 84.61   \\
				&Proposed w/o LDA         & 58.69          &80.89      & 68.03 & 75.20  & 70.74          &85.55      & 77.44  & 82.11\\
                &Proposed w/o LDA, $L_{SDF}$     & 59.98           & 77.17      & 62.12 & 70.35 & 69.89           & 81.36      & 75.19 & 78.77 \\\bottomrule
\end{tabular}}
\label{ablations}
\end{table*}

\subsection{Comparison with other state-of-the-art segmentation methods}
In this section, we compare our proposed method with other state-of-the-art (SOTA) methods for TLS detection. The UNet\cite{2015U}, Deeplab v3+\cite{2018Encoder}, and  nnUNet \cite{2018nnU} are included in our experiments for their excellent performance in medical image segmentation. All these methods are trained on the  NDTH training set  and evaluated on the NDTH validation set and the JHCM dataset. Quantitative comparison results are presented on Table \ref{com_sota}.

\begin{figure*}[htbp]
		\centering
		{
			\includegraphics[width=15cm,]{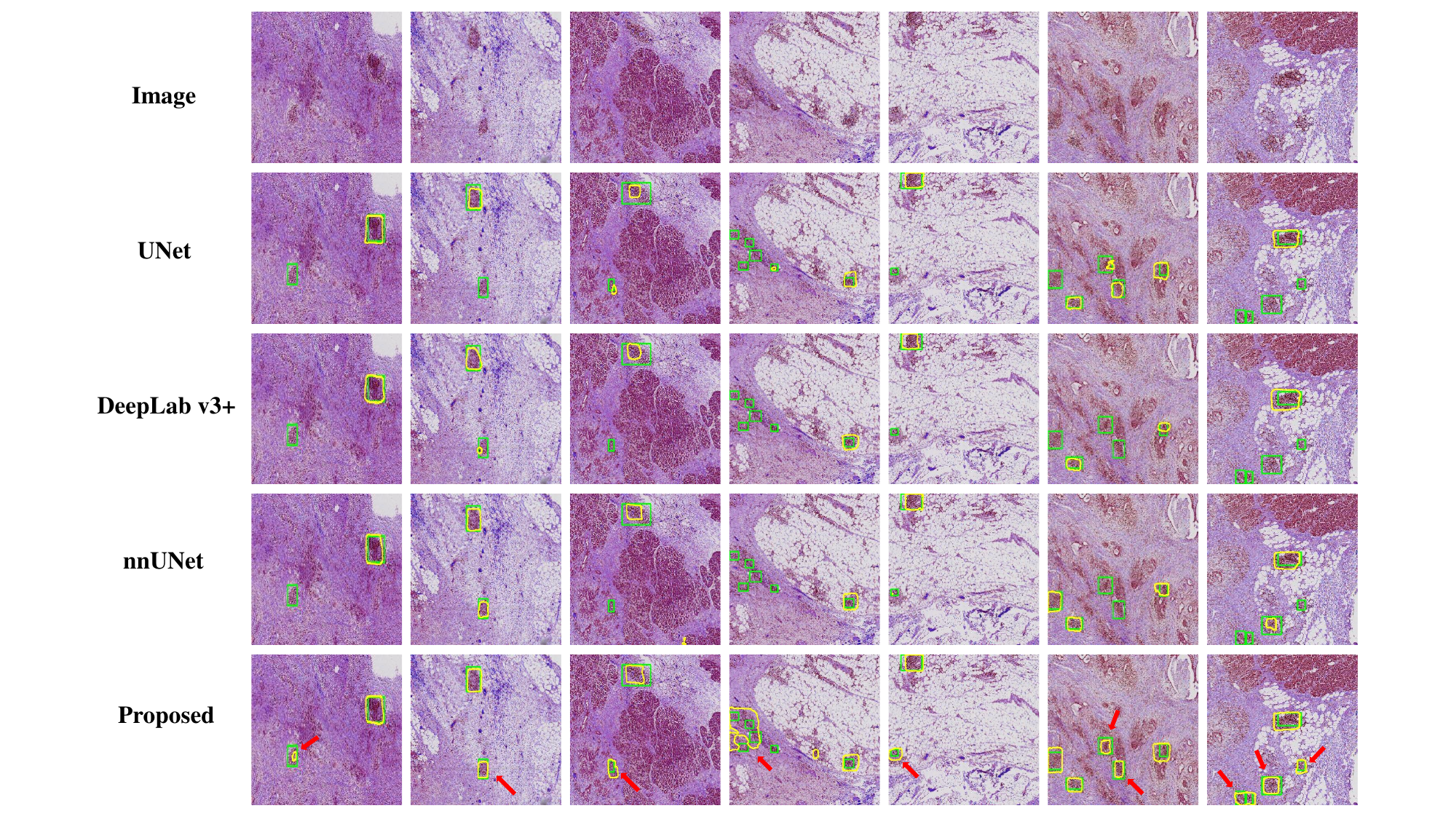}
		}
		\caption{Visual comparison results between the SOTA segmentation methods and our proposed method. The green and yellow denote the ground truth and predictions, respectively. Improved predictions are signed by red arrows.}\label{cp_vis}
	\end{figure*}

Experimental results in Table \ref{com_sota} demonstrate that our proposed method significantly outperforms these SOTA methods, and achieves the best performance in recall, $F_1$ score, $F_2$ score, SP, $GF_1$ score and $GF_2$ score. We observe that the generic UNet \cite{2015U} and deeplab v3+ \cite{2018Encoder} always have conservative predictions by minimizing the false positive predictions as much as possible, despite causing many false negative predictions. Therefore, they have a higher precision and SP than ours, but leading to  18.07\%-44.36\%, 6.58\%-21.01\%,  13.68\%-35.25\%, 20.48\%-47.35\%, 9.25\%-29.10\% and 16.28\%-40.68\% decrease in recall, $F_1$ score, $F_2$ score, $GF_1$ score and $GF_2$ score, respectively comparing with ours. Figure \ref{cp_vis} shows the visual comparison of TLS detection results between the above methods and ours.

\begin{table*}[htbp]
\caption{Quantitative comparison results with other SOTA methods for TLSs detection on two collected dataset. Experimental results on the NDTH validation set and the JHCM dataset are presented. P, R, F1 and F2 denote the precision, recall, $F_1$ score and $F_2$ score, respectively. SP, BR, GF1 and GF2 represent the introduced segmentation precision, box recall, $GF_1$ score and $GF_2$ score, respectively.}
\setlength{\tabcolsep}{4.8mm}{}{
\begin{tabular}{clcccccccc}
\toprule
Datasets                & Methods             & P     & R        & F1  &F2     &  SP & BR & GF1  & GF2        \\ \midrule
\multirow{4}{*}{NDTH}   & UNet\cite{2015U}                 & \textbf{85.27} & 66.27          & 74.58  & 69.36     & \textbf{89.92} & 67.47          & 77.09 & 71.02  \\
                        & DeepLab v3+\cite{2018Encoder}           & 84.68          & 63.25          & 72.41 & 66.62   & 85.48          & 63.86          & 73.10 &67.26     \\
                        & nnUNet\cite{2018nnU}                & 77.92          & 72.29          & 75.00  & 73.35  & 86.21          & 74.70          & 80.04 &76.75      \\
                        & Proposed    & 78.21          & \textbf{84.34} & \textbf{81.16} & \textbf{83.04}  & 84.80          & \textbf{87.95} & \textbf{86.34} &\textbf{87.30}\\ \midrule
\multirow{4}{*}{JHCM} & UNet\cite{2015U}                & \textbf{76.57}          & 45.39          & 56.99  & 49.11   & \textbf{82.88}          & 49.02          & 61.60   & 53.38   \\
                        & DeepLab v3+\cite{2018Encoder}          & 71.11          & 38.77          & 50.18 & 42.65  & 74.78          & 40.35          & 52.42      &  44.44   \\
                        & nnUNet\cite{2018nnU}                  &59.98                &77.17                & 62.12     & 70.35  &69.89                &81.36                & 75.19    &  78.77     \\
                        & Proposed               & 62.25               & \textbf{83.13}               & \textbf{71.19}    & \textbf{77.90}   & 76.15               & \textbf{87.70}               & \textbf{81.52}  & \textbf{85.12}    \\ \bottomrule
\end{tabular}}
\label{com_sota}
\end{table*}

\subsection{Application for studying the relationship between TLS density and peripancreatic vascular invasion}

For patients with pancreatic tumors accompanied by peripancreatic vascular invasion, especially venous invasion, whether the peripancreatic vessel is invaded determines the direction of treatment, and also affect the survival time and prognosis of patients \cite{1996Vascular, 2008A, 2016Preoperative}. In this section, we apply our proposed TLS detection method to study the relationship between TLS density and peripancreatic vascular invasion.

We collect the clinical information from 12 patients in NDTH dataset and 40 patients (there is a patient whose clinical information is not available) in JHCM datatset to carry out our statistical experiments. 
There are 19 patients without peripancreatic vascular invasion and 33 patients with peripancreatic vascular invasion. Therefore, we divide the 52 samples into two groups: the no-invasion group and the invasion group. Then we apply our proposed method to detect the TLSs on their WSIs, and calculate the TLS density of the patients in the two groups, respectively. The TLS density is calculated as the ratio of the TLS number to the WSI's area.


The Shapiro-Wilk (S-W) test \cite{1965An} is conducted to check whether the TLS densities of the 52 patients obey the normal distribution. Based on the S-W test results of $p<0.05$ and $p<0.001$ on the invasion and no-invasion group, respectively (i.e., they do not obey the normal distribution), we use the Mann-Whitney U test \cite{1947On} to observe the correlation between the TLS density and the peripancreatic vascular invasion. We get the result of $p=0.03$ for the Mann-Whitney U test, which indicates that the TLS density  is indeed related to the peripancreatic vascular invasion.


For comparison, we also study on the correlation between the lymphocyte density and the peripancreatic vascular invasion. The same statistical testing experiments are conducted for the lymphocyte density. We get the S-W test results of $p<0.01$ on the invasion group (i.e., they do not obey the normal distribution). We use the Mann-Whitney U test to observe the correlation between the TLS density and the peripancreatic vascular invasion. The we get the result of $p=0.1$ for the Mann-Whitney U test, which indicates that the lymphocyte density has no significant relationship to the peripancreatic vascular invasion. Figure \ref{box} presents the distributions of the TLS density and lymphocyte density for the two groups.

\begin{figure*}[htbp]\label{box}
		\centering
		{
			\includegraphics[width=15cm,]{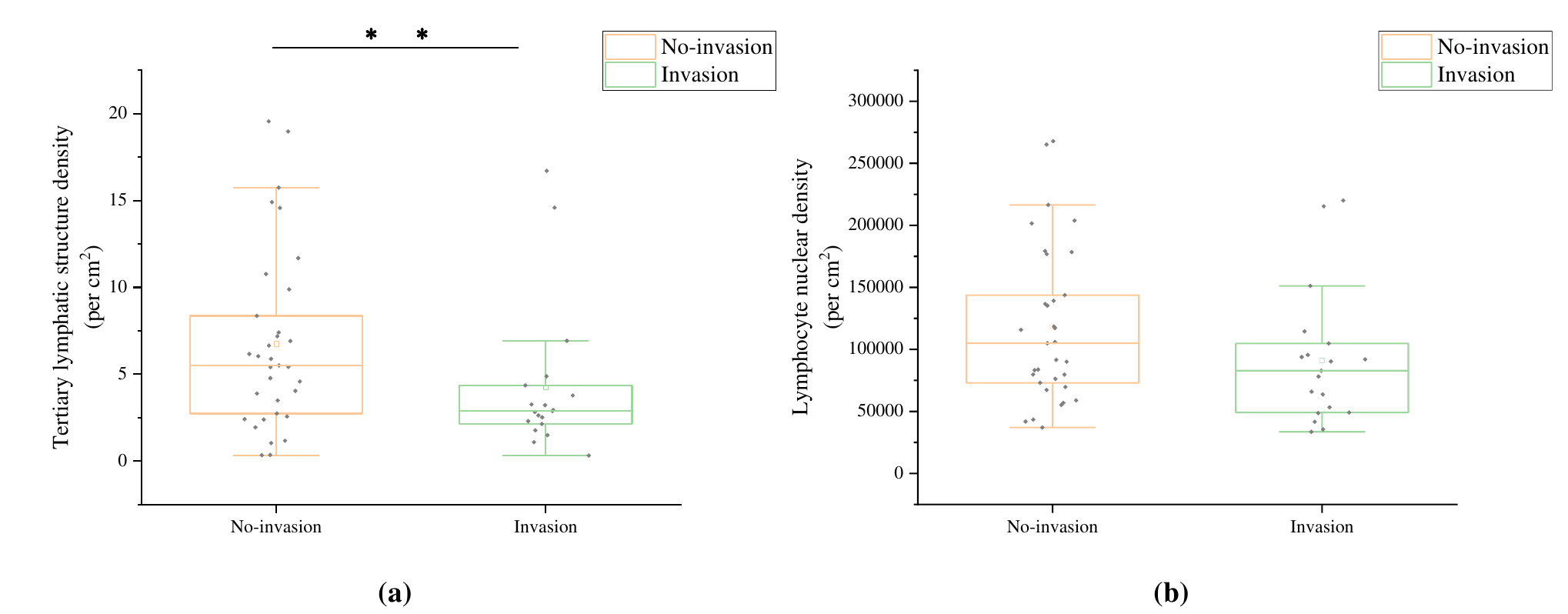}
		}
		\caption{The distribution of (a) TLS density and (b) lymphocyte density for two groups (the no-invasion group and the invasion group). $**$ indicates $p < 0.05$ using the Mann-Whitney U test.}\label{box}
	\end{figure*}

\section{Discussions}
In this paper, we develop a novel weakly-supervised segmentation framework for TLS detection. Since TLSs are aggregates of lymphocytes, we utilize the density map of lymphocyte nuclei as an attention guidance in our approach. Although the segmentation of individual lymphocyte nucleus may not affect the TLS detection, the overall accuracy of lymphocyte nuclei segmentation results may still impact the detection results. Therefore, we also show the lymphocyte nuclei segmentation results in this section. As described in Section \ref{L-seg}, we have used a domain-adversarial approach to segment lymphocyte nuclei on pancreatic pathology images.

\begin{figure*}[htbp]
		\centering
		{
			\includegraphics[width=12cm,]{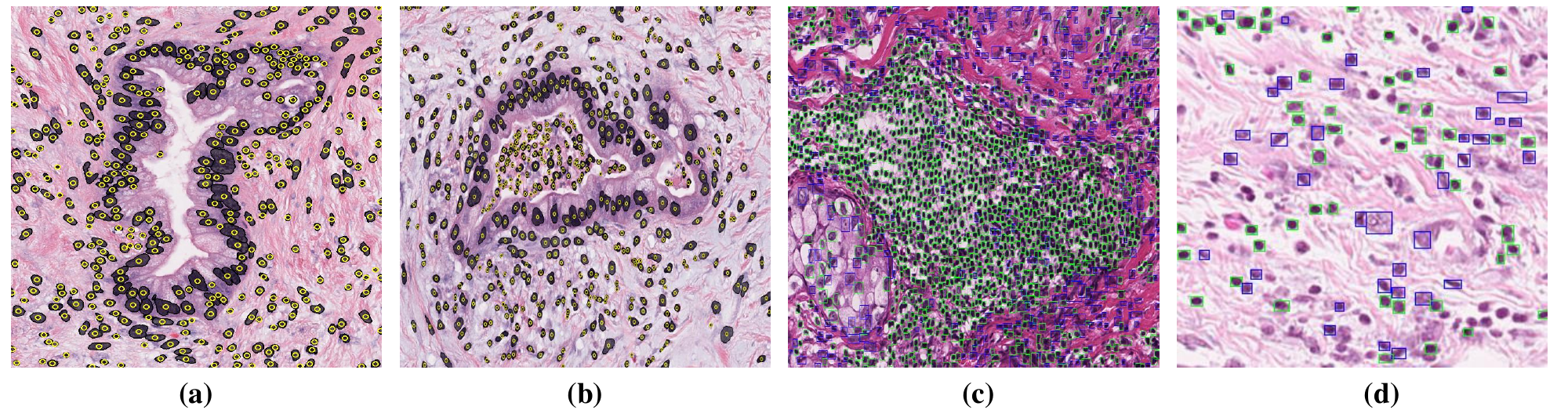}
		}
		\caption{Segmentation and classification results of the designed combination of baseline model for nuclei segmentation and domain adversarial network for lymphocyte nuclei recognition. (a)-(b): Segmentation results. The predicted nuclei are shown in dark masks whose centers are marked by yellow circles. (c)-(d): Classification results. The lymphocyte and non-lymphocyte nuclei are shown in green and blue boxes, respectively.}\label{HD_vis}
	\end{figure*}

We show the segmentation and classification results of our designed combination of the baseline model for nuclei segmentation and the domain adversarial network for lymphocyte nuclei recognition in Table \ref{seg_results}. The segmentation accuracy, sensitivity, and specificity for the nucleus achieve 86.20\%, 76.29\%, and 80.94\%, respectively. The classification accuracy, sensitivity, and specificity for lymphocyte and non-lymphocyte nuclei achieve 88.87\%, 94.79\%, and 34.41\%, respectively. Figure \ref{HD_vis} presents two visual results for the segmentation and classification, respectively.

\begin{table}[htbp]
        \centering
		\caption{Segmentation and classification results for lymphocyte nucleus on our test dataset.}
		\vspace{0pt}	
		\renewcommand\arraystretch{2}
		\setlength{\tabcolsep}{4mm}{}{
			\begin{tabular}{lccc}
				\hline
				  Methods                     &Accuracy       & Sensitivity      & Specificity    \\
				\hline
                  HoVerNet                   &86.20                   &76.29               &80.94    \\
				DANN     &88.87                &94.79                     &34.41         \\
				\hline
		\end{tabular}}
		\label{seg_results}
	\end{table}

We would like to highlight that we have opted for a simple U-shape segmentation network to detect TLSs rather than a complex object detection framework. This decision is based on the fact that detection frameworks usually require a larger amount of annotated data and parameters to be trained compared to the U-shape segmentation network. Therefore, in the context of few-shot learning, the detection framework does not notably outperform the U-shape network. It is worth noting that the designed combination of the baseline model for nuclei segmentation and domain adversarial network for lymphocyte nuclei classification can still be improved, especially in terms of lymphocyte nuclei recognition specificity.


The proposed LDA improves the TLS segmentation performance by multi-channel fusion. However, the proposed LDA can also be embedded in the U-shape network architecture or used as an extra loss function for training, which can be studied in our subsequent work. Additionally, we set the patch size $d$ for coarse-scale pooling and fine-scale density calculation as 32 based on medical experts' experience. 

Our proposed LDA shows an improvement in TLS segmentation performance through multi-channel fusion. However, we intend to investigate further the potential benefits of embedding the proposed LDA in the U-shape network architecture or using it as an additional loss function during training in future work. Furthermore, we have chosen a patch size $d$ of 32 for coarse-scale pooling and fine-scale density calculation based on medical experts' recommendations.

Regarding the detected TLSs, we have examined the relationship between TLSs and vascular invasion. Through a study of two central patients, we found that TLSs have a stronger association with vascular invasion than individual lymphocytes, highlighting TLSs as an important component of the tumor immune microenvironment. With the expansion of sample size and the subdivision of lymphocyte categories, we will conduct more detailed and in-depth investigations of the immune microenvironment.

\section{Conclusions}
In this paper, we propose a novel weakly supervised segmentation network embedding cross-scale attention guidance and noise-sensitive constraint for TLS detection. We firstly obtain the segmentation and classification results of the lymphocyte nuclei by combining a pretrained baseline model for nuclei segmentation and a domain adversarial network for lymphocyte nuclei recognition. Then, we establish a cross-scale attention guidance network by jointly learning the coarse-scale features from the original H\&E images and fine-scale features from our calculated lymphocyte density attention. A noise-sensitive constraint is introduced by embedding the signed distance function loss in the training procedure to reduce tiny segmentation errors. Experimental results on two collected datasets demonstrate that our proposed algorithm outperforms the state-of-the-art segmentation-based methods for TLS detection. Additionally, we apply our method to validate that the TLS density is significantly related to the peripancreatic vascular invasion based on the clinical data acquired from two independent institutions. Our proposed approach can be applied to the TLS analysis for the tumors in other organs.

\section*{Acknowledgment}
This work is supported by China's Ministry of Science and Technology(No. SQ2020YFA0713800), the National Natural Science Foundation of China(No. 11971229, 12001273), and the Fundamental Research Funds for the Central Universities(No.30922010904).

\ifCLASSOPTIONcaptionsoff
  \newpage
\fi



%

\bibliographystyle{ieeetr}
\bibliography{TLSs-bibliography.bib}%

%








\end{document}